\documentclass[aps,prl,twocolumn,superscriptaddress]{revtex4-1}

\usepackage{graphicx}
\usepackage{natbib}
\usepackage[breaklinks]{hyperref}
\newcommand{\ket}[1]{|{#1}\rangle}

\newcommand{\NID}{\text{NID}}
\newcommand{\NCD}{\text{NCD}}

\begin{document}

\title{Probing quantum-classical boundary with compression software}
\author{Hoh~Shun~Poh}
\affiliation{Center~for~Quantum~Technologies, National~University~of~Singapore,
3 Science Drive 2, Singapore  117543}
\author{Marcin~Markiewicz}
\affiliation{Faculty~of~Physics, University~of~Warsaw,  ul. Pasteura 5, PL-02-093
Warszawa, Poland}
\affiliation{Institute~of~Theoretical~Physics~and~Astrophysics, University~of~Gdansk, ul. Wita Stwosza 57, PL-80-952, Gdansk, Poland}
\author{Pawe{\l}~Kurzy\'nski}
\affiliation{Center~for~Quantum~Technologies, National~University~of~Singapore, 3 Science Drive 2, Singapore  117543}
\affiliation{Faculty~of~Physics, Adam~Mickiewicz~University, Umultowska 85, 61-614 Pozna\'{n}, Poland}
\author{Alessandro~Cer\`{e}}
\affiliation{Center~for~Quantum~Technologies, National~University~of~Singapore, 3 Science Drive 2, Singapore  117543}
\author{Dagomir~Kaszlikowski}
\affiliation{Center~for~Quantum~Technologies, National~University~of~Singapore,
3 Science Drive 2, Singapore  117543}
\affiliation{Department~of~Physics,
  National~University~of~Singapore, 2 Science Drive 3, Singapore  117542}
\author{Christian~Kurtsiefer}
\affiliation{Center~for~Quantum~Technologies, National~University~of~Singapore,
3 Science Drive 2, Singapore  117543}
\affiliation{Department~of~Physics,
  National~University~of~Singapore, 2 Science Drive 3, Singapore  117542}
\begin{abstract}
We experimentally demonstrate that it is impossible to simulate quantum
bipartite correlations with a deterministic universal Turing machine. Our
approach is based on the \emph{Normalized Information
  Distance} (NID) that allows the comparison of two pieces of data without
detailed knowledge about their origin. Using NID, we derive an inequality for
output of 
two local deterministic universal Turing
machines with correlated inputs. This inequality is violated by correlations
generated by a maximally entangled polarization state of two photons. The
violation is shown using
a freely available lossless compression program. The presented technique may
allow to complement the common statistical interpretation of quantum physics
by an algorithmic one.
\end{abstract}

\pacs{
03.67.-a, 	
03.65.Ta, 	
42.50.Dv, 	
89.20.Ff 	
}

\maketitle

\section{Introduction}

The idea that physical processes can be considered as computations done on
some universal machines traces back to Turing and von
Neumann~\cite{Coveney:1996tx},
and the growth of the computational power allowed for further development of these concepts.
This resulted in a completely new approach to science in which the complexity of observed phenomena is closely related to the complexity of computational resources needed to simulate them~\cite{Wolfram:1985je}.
In addition, there are physical phenomena that simply cannot be traced with analytical tools,
which further motivated a computational approach to physics~\cite{Moore:1990bn}.
Moreover, the idea of quantum computation~\cite{Feynman:82} lead to a
discovery of a few problems that seem not
efficiently traceable on classical computers but efficiently
on a quantum version~\cite{Shor:1994tb,Grover:1996}.

Classical physics can be simulated on universal Turing machines, or other
computationally equivalent models~\cite{Wolfram:2001ue}.  On the other hand,
efficient simulation of quantum systems requires a replacement of
deterministic universal Turing machines with 
quantum computers whose states are non-classically correlated. Such machines
can even simulate any local quantum system efficiently~\cite{Deutsch:1985ba,Lloyd:96}.
Can we experimentally distinguish between these two descriptions of the universe
using a logically self-contained computational approach?

In this paper, we show that
there are processes which cannot be simulated on local classical machines at all,
independently of the available classical resources.
We first introduce the notion of Kolmogorov complexity,
a measure of the classical complexity of a phenomena,
and later apply it to derive a bound on classical descriptions~\cite{Li:2004kz}. Next, we use the fact that Kolmogorov complexity can be approximated by compression algorithms~\cite{Cilibrasi:2005jna}.

We compress
experimental data obtained from polarisation
measurements on entangled photon pairs
and show the violation of a classical bound.

Let's consider the description of a machine, whether classical or quantum, that outputs a string $x$ made of 0's and 1's.
In the case of a Turing machine $U$, we can always write a program $\Lambda$ that generates $x$.
The simplest such program is obviously `PRINT $x$'.
However, this is not optimal: in many cases the program can be much shorter than the string itself.

This brings us to the concept of Kolmogorov complexity $K(x)$, the minimal
length $l(\Lambda)$ of all programs $\Lambda$ that reproduce a specific output
$x$. If $K(x)$ is of the order of
the length of the output
$l(x)$ then our algorithmic description of $x$ is inefficient, and $x$ is called algorithmically random~\cite{Li:2009wu}.
In most
cases $K(x)$ cannot be computed~\cite{Cover:2006ub}.
To circumvent this issue, we can estimate $K(x)$ with some efficient lossless compression algorithm $C(x)$~\cite{Cilibrasi:2005jna}.

We now extend this picture by considering two Turing machines
$U_A$ (Alice) and $U_B$ (Bob), which are spatially separated.
If these machines cannot communicate,
they generate two output strings that are independent,
although the programs fed into the machines can be correlated.
Moreover, the input programs are classical bit strings so the correlations
between them must be classical.

We determine the complexity of the generated strings using the
\emph{Normalized Information Distance} (NID)~\cite{Li:2004kz}.
This distance allows for a comparison of two data sets without detailed knowledge about their origin.
In practice, we evaluate an approximation to the NID, the \emph{Normalized Compression Distance} (NCD)~\cite{Cilibrasi:2005jna}, using a lossless compression software, in our case the LZMA Utilities, based on the Lempel-Ziv-Markov chain algorithm~\cite{ref:lzma}.

\section{Simulation by deterministic universal Turing machines}
\begin{figure}
    \begin{center}
        \includegraphics[width=.8\columnwidth]{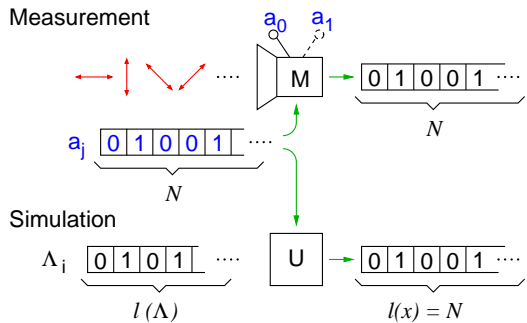}
        \caption{Measurement: $N$ particles enter a measuring device,
          characterized by two possible polarizer settings $a_0$ and $a_1$,
          which generates a bit string of $N$ measurement outcomes.
          Simulation: a universal Turing machine is fed with a
          program elements $\Lambda_i$ and information about the chosen setting
          $a_0$ or $a_1$. It delivers an output string of length $N$.}
        \label{fig0}
    \end{center}
\end{figure}
We
consider a model experiment, similar to the one used for testing the Bell
inequalities~\cite{Bell}: a source emits pairs of photons that travel to two
separate polarization analyzers M$_A$ (Alice) and M$_B$ (Bob).
Each analyzer has two outputs associated with bit values 0 and 1,
and can be
set along directions $a_0$ or $a_1$ for M$_A$, and $b_0$ or $b_1$ for M$_B$.
The record of the outputs from each analyzer forms a bit string (see Fig.~\ref{fig0}).

The output $x$ of
each individual analyzer can be described as the output of a Turing machine $U$, fed with the settings $a_j$ or $b_k$, and a program $\Lambda$.
The program will contain the information for
generating the correct output for
every detection event and for every setting.

If we consider a string of finite length $l(x)=N$, $\Lambda$ will have to describe the $4^N$ possible events.
The length of the shortest $\Lambda$ is equal to the Kolmogorov complexity of the generated string.

\begin{figure}
    \begin{center}
        \includegraphics[width=.8\columnwidth]{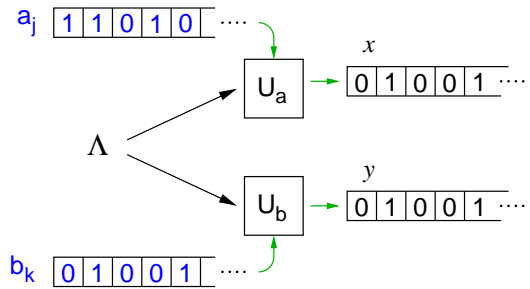}
        \caption{Local classical machines simulating the generation of strings $x$ and $y$ by correlation measurements on an entangled state.}
        \label{fig:turing_pair}
    \end{center}
\end{figure}

Next, we consider the simulation of the experiment with two local non-communicating machines $U_A$ and $U_B$ (see Fig.~\ref{fig:turing_pair}).
We feed a program $\Lambda$ to
both of them and obtain two output strings, $x$ and $y$, both of length $N$.
In this case, the program has to describe the behavior of all $2N$
events
for all possible settings $a_j$ and $b_k$, hence $16^N$ possible events.
\subsection{Normalized Information Distance}
The Kolmogorov complexity of two bit strings $K(x, y)$ is the length of the shortest program generating them simultaneously. $K(x, y)$ can be shorter than $K(x)\,+\,K(y)$ if $x$ and $y$ are correlated - the more correlated they are, the simpler it is to compute one string knowing the other. This idea was further carried out by Cilibrasi and Vitanyi~\cite{Cilibrasi:2005jna} who constructed a distance measure between $x$ and $y$ called \emph{Normalized Information Distance} (NID),
\begin{equation}
    \NID(x, y)\,=\,\frac{K(x,y)\,-\,\min\{K(x), K(y)\}}{\max\{K(x), K(y)\}}\,.
\end{equation}

The NID obeys all required properties of a metric, in particular, the triangle inequality
\begin{equation}\label{eq:triangle}
    \NID(x, y)\,+\,\NID(y, z)\,\geq\,\NID(x, z)\,.
\end{equation}
The above inequality holds up to a correction of order $\log(l(x))$, which can
be neglected for sufficiently long strings~\cite{Cilibrasi:2005jna}.

\subsection{Information Inequality}
We consider the bit strings $x_{a_j}$ and $y_{b_k}$ generated by Alice and Bob
with fixed setting $a_j$ and $b_k$.
Equation~(\ref{eq:triangle}) then transforms into 
\begin{equation}\label{eq:NID_a0b1}
    \NID(x_{a_0}, y_{b_0})+\NID(y_{b_0}, y_{b_1})\geq\NID(x_{a_0}, y_{b_1})\,.
\end{equation}
However, $\NID(y_{b_0}, y_{b_1})$ cannot be determined experimentally because the strings $y_{b_0}$ and $ y_{b_1}$ come from measurements of incompatible observables.
We therefore use the triangle inequality
\begin{equation}\label{eq:NID_b0b1}
    \NID(x_{a_1}, y_{b_0})+\NID(x_{a_1}, y_{b_1})\geq\NID(y_{b_0}, y_{b_1})\,,
\end{equation}
and combine it with inequality~(\ref{eq:NID_a0b1})
to obtain a quadrangle inequality:
\begin{eqnarray}\label{eq:ineq}
    \NID(x_{a_0}, y_{b_0})+\NID(x_{a_1}, y_{b_0}) + &\NID(x_{a_1}, y_{b_1})\geq \nonumber\\
    &\NID(x_{a_0}, y_{b_1}).
\end{eqnarray}

Similar to various tests of Bell inequalities, we introduce a scalar quantity
$S'$ that quantifies the degree of violation of Eq.~(\ref{eq:ineq}):
\begin{eqnarray}\label{eq:ineq2}
   S' &=& \NID(x_{a_0}, y_{b_1}) -  \NID(x_{a_0},y_{b_0}) \nonumber \\
   &-& \NID(x_{a_1},y_{b_0})\,-\,\NID(x_{a_1},y_{b_1}) \leq 0
\end{eqnarray}

In order to experimentally test this inequality,
we have to address the following problem.
We can set up a source to generate entangled photon pairs in a state of our
choosing, but
we cannot control the nature of the measurement.
For every experimental run $i$ with the same preparation the resulting string $x_{i,a_j}$ can be different.
Consequently, the corresponding program $\Lambda_i$ is different for every experimental run.

It is reasonable to assume that for every two experimental runs $i$ and $i'$ the complexity of the generated strings  remains the same:
$K(x_{i,a_j})\,=\,K(x_{i',a_j})$ and $K(x_{i,a_j}, y_{i,b_k})\,=\,K(x_{i',a_j}, y_{i',b_k})$.
Without these assumptions
the same physical preparation of the experiment has different consequences and thus the notion of preparation loses its meaning. More generally, the predictive power of science can be expressed by saying that the same preparation results in the same complexity of observed phenomena.

\section{Estimation of Kolmogorov complexity}

In general the Kolmogorov complexity cannot be evaluated, but it can be estimated. One can adapt two conceptually different approaches.

\subsection{Statistical Approach}

This 
approach takes into account the ensemble of all possible $N$-bit strings and asks
about their average Kolmogorov complexity. It can be shown that this average
equals  the Shannon entropy $H(X)$ of the ensemble~\cite{Cover:2006ub}, and thus
\begin{equation}\label{eq:nid_entropy}
    \langle \NID(x, y)\rangle=\frac{H(x, y)-\min\{H(x), H(y)\}}{\max\{H(x), H(y)\}}\,.
\end{equation}

Inequality (\ref{eq:ineq})
becomes a type of entropic Bell inequality introduced by Braunstein and
Caves~\cite{Braunstein:1988en} if local entropies are maximal, i.e.,
$H(x)=H(y)=N$. 
They showed that for a maximally entangled polarization state of two photons,
and polarizer angles obeying the constraints
\begin{eqnarray}
 \vec{a_0}\cdot\vec{b_1}&=&\cos 3\theta\,,\nonumber \\
 \vec{a_0}\cdot\vec{b_0}&=&\vec{a_1}\cdot\vec{b_0}=\vec{a_1}\cdot\vec{b_1}=\cos
 \theta\,,\label{eq:braunstein}
\end{eqnarray}
inequality~(\ref{eq:ineq}) is violated for an appropriate range of $\theta$.
Calculating the entropy $H(x, y)$ using
the probability distributions predicted by quantum mechanics,
it is possible to obtain the expected value of $S'$ as a function of $\theta$
(Fig.~\ref{fig:violation_versus_separation}a).
The maximal violation of this inequality is $S'=0.24$, with a separation of
$\theta\,=\,8.6^\circ$.

\subsection{Algorithmic approach}
It is possible
to avoid a statistical description of our experiment
following the ideas pioneered in~\cite{Cilibrasi:2005jna}. There, it was shown that the Kolmogorov complexity can be well approximated by
the application of 
compression algorithms.
This approximation introduces the new distance called \emph{Normalized Compression Distance}~(NCD)
\begin{equation}\label{eq:NCD}
    \NCD(x,y)=\frac{C(x,y)-\min\{C(x),C(y)\}}{\max\{C(x),C(y)\}}\,,
\end{equation}
where $C(x)$ is the length of the compressed string $x$, and $C(x,y)$ is the length of the compressed concatenated strings $x,y$.
Replacing NID with NCD in Eq.~(\ref{eq:ineq2}) leads to a new inequality:
\begin{eqnarray}\label{eq:ineq3}
   S'\rightarrow S&=&\NCD(x_{a_0},y_{b_1}) - \NCD(x_{a_0},y_{b_0}) \nonumber
   \\ &-&\NCD(x_{a_1},y_{b_0}) - \NCD(x_{a_1},y_{b_0}) \leq 0\,.
\end{eqnarray}
This expression can be tested experimentally
because the NCD distance measure is operationally defined.

\section{Choice of compressor}

Before moving to the experiment, we need to ensure the suitability of the
compression software we use to evaluate the NCD. For this, we 
numerically simulate the outcome of an experiment, based on a distribution of
results predicted by quantum physics.
Among the packages we tested, we found that the LZMA Utility~\cite{ref:lzma}
approaches the Shannon limit~\cite{Shannon:1948iy} most closely.

The simulation also allows us to verify the angle that maximizes the
violation of Eq.~(\ref{eq:ineq3}) predicted from Eq.~(\ref{eq:braunstein}).
The results of the simulation are presented in Fig.~\ref{fig:violation_versus_separation}.
More details on the generation of the simulated data and the choice of the compressor are provided in the Appendix.

\section{Experiment}

\begin{figure}
    \begin{center}
        \includegraphics[width = 0.9\columnwidth]{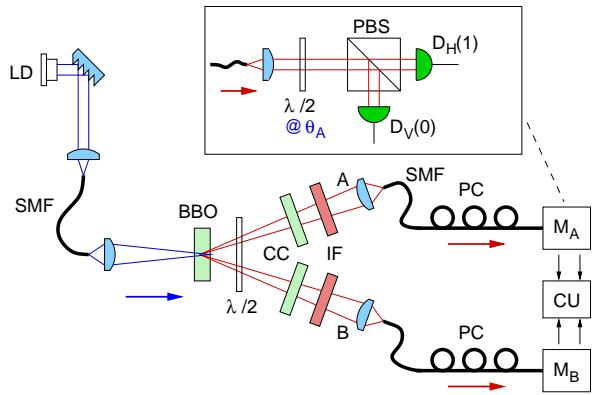}
        \caption{Schematic of the experimental set-up. 
        Polarization correlations of entangled-photon pairs are measured by the polarization analyzers
        M$_A$ and M$_B$,
        each consisting of a half wave plate ($\lambda/2$) followed by a polarization beam splitter (PBS).
        All photons are detected by Avalanche photodetectors D$_H$ and D$_V$, and registered in a coincidence unit (CU).}
        \label{fig:setup}
    \end{center}
\end{figure}

In our experiment (see Fig.~\ref{fig:setup}), the output of a
grating-stabilized laser diode (LD, central wavelength 405\,nm) passes
through a single mode optical fiber (SMF) for spatial mode filtering, and is
focused to a beam waist of 80\,${\mu}$m into a 2\,mm thick BBO crystal. In
this crystal (cut for type-II phase-matching), photon pairs are generated via
spontaneous parametric down-conversion (SPDC) in a slightly non-collinear
configuration.  A half-wave plate ($\lambda$/2) and a pair of compensation
crystals (CC) take care of the temporal and transversal
walk-off~\cite{kwiat:95}.
Two spatial modes (labeled A and B) of down-converted light, defined
by the SMFs for 810\,nm, are matched to the pump mode to optimize
the
collection~\cite{kurtsiefer:01}. In type-II SPDC, each down-converted pair
consists of an ordinary and extraordinarily polarized photon, corresponding to
horizontal (H) and vertical (V) in our setup.  A pair of
polarization controllers (PC) ensures that the SMFs do not affect the
polarization of the collected photons. To arrive at an approximate singlet
Bell state, the phase $45^\circ$ between the two decay possibilities in the polarization state
\begin{equation}
    \ket{\psi}={1\over\sqrt{2}}\left(\ket{H}_A\ket{V}_B+e^{i45^\circ}\ket{V}\ket{H}_B\right),
    \label{eq:gen_ouput_state}
\end{equation}
is adjusted to $45^\circ$\,=\,$\pi$ by tilting the CC.

In the polarization analyzers (Fig.~\ref{fig:setup}), the photons from SPDC
are projected onto arbitrary linear polarization by $\lambda$/2 plates, set
to half of the analyzing angles $\theta_{A(B)}$, and
polarization beam splitter (PBS) in each analyzer.
Photons are detected by avalanche photo diodes (APDs), and corresponding detection events from the same pair identified by a coincidence unit (CU) if they arrive within $\approx\pm$3\,ns of each other.

The quality of polarization entanglement is tested by probing the polarization
correlations in a basis complementary to the intrinsic HV basis of the
crystal; for Bell states $\ket{\psi^\pm}$, strong polarization correlations
are e.g. expected in a $\pm$\,45$^\circ$ linear polarization basis.

With interference filters (IF) of 5\,nm bandwidth (FWHM) centered at 810\,nm,
we observe a visibility $V_{45}$ = 99.9$\pm$0.1\%. The visibility in
the natural H/V basis of the type-II down-conversion process also reaches
$V_{\rm HV}$ = 99.9$\pm$0.1\%. A separate test of a CHSH-type Bell inequality~\cite{clauser:69}
leads to a value of $S=2.826\pm0.0015$. This indicates a relatively high quality
of polarization entanglement; the uncertainties in the visibilities are obtained from propagated Poissonian counting statistics.

\subsection{Measurement and Data Post-processing}

We record two-fold coincidences of detection events between
detectors at A and B.
For each PBS, the transmitted output is associated with 0 and the reflected one with 1.
The
resulting binary strings $x$ from A, and $y$ from B are written into two
individual binary files.
From these, we calculate the NCD using Eq.~(\ref{eq:NCD}).
This procedure is repeated for each of the four settings ($a_0,b_0$), ($a_1,b_0$), ($a_1,b_1$), and ($a_0,b_1$) in order to obtain the value for $S$.

To remove the bias due to differences in the detection efficiency of the APDs
in the experiment, we also measure for each setting the associated orthogonal
ones (see Appendix for details).

\begin{figure}
    \begin{center}
        \includegraphics[width = 0.95\columnwidth]{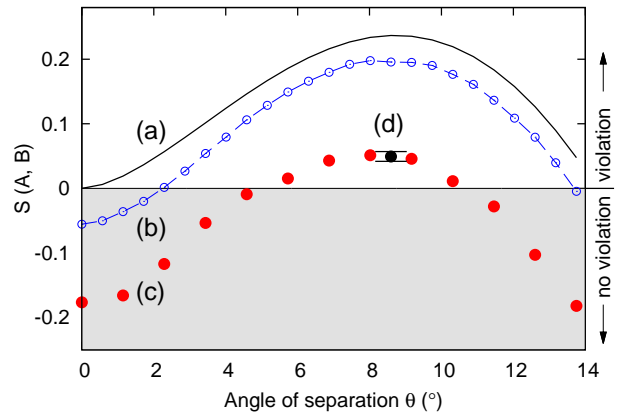}
        \caption{\label{fig:violation_versus_separation}
        Plots of $S$ versus angle of separation $\theta$.
        (a) Result obtained from Eq.~(\ref{eq:nid_entropy})
        (b) result obtained from using the LZMA compressor on a simulated data ensemble,
        (c) measurement of $S$ in the experiment shown in Fig.~\ref{fig:setup},
        and (d) longer measurement at the optimal angle $\theta=8.6^\circ$.}
    \end{center}
\end{figure}

\section{Results}

The inequality is experimentally tested by evaluating $S$ in
Eq.~(\ref{eq:ineq3}) for a range of $\theta$; the
obtained values [points (c), (d) in Fig.~\ref{fig:violation_versus_separation}]
are
consistently lower than the trace (a) calculated via entropy
using Eq.~(\ref{eq:nid_entropy}),
and than a simulation with the same compressor (b).
This is because the LZMA Utility is not working exactly at the Shannon limit,
and also due to imperfect state generation and detection.

As a consequence of Eq.~(\ref{eq:braunstein}), we expect the maximal violation for $\theta=8.6^\circ$.
For this particular angle we collected results from a large number of photon
pairs. Although we set out in this work to avoid a statistical argument in the
interpretation of measurement results, we do resort to statistical techniques to
assess the confidence in an experimental finding of a violation of inequality
Eq.~(\ref{eq:ineq3}).
To estimate an uncertainty of the experimentally obtained values for $S$,
this large data set
was subdivided
into files with length greater than $10^5$ bits.
The results from all the subdivided files are then averaged to obtain the
final result of $S(\theta=8.6^\circ)$=\,0.0494$\,\pm\,$0.0076, with the latter
indicating a relatively small standard deviation over these different subsets.

\section{Conclusion}
There is a trend to look at physical systems and processes as
programs run on a computer made of the constituents of our universe. We could show
that this is not possible if one uses a computation paradigm of a local
deterministic Turing machine. Although this has been already extensively
researched in quantum information theory, we present a complementary
algorithmic approach for an explicit, experimentally testable example.
This algorithmic approach is complementary to the orthodox Bell inequality approach to quantum nonlocality~\cite{Bell} that is statistical in its nature.

Any process that can be simulated on a local universal Turing machine can be encoded as a program that is fed into it. For every such a program there exists its shortest description called Kolmogorov complexity, which in most of the cases can only be  approximated using compression software. Moreover, such a description must obey distance properties as shown in~\cite{Li:2004kz,Cilibrasi:2005jna}.
By testing Eq.~(\ref{eq:ineq3}), we showed that this is unattainable in the
specific case of polarization-entangled photon pairs. Therefore, there
exist physical processes that cannot be simulated on local universal Turing
machines.

There are two fundamentally different notions of complexity in computer
science. On one hand, computational complexity, mainly researched on in quantum information science, studies how much resources are needed to solve a computational problem. These studies focus on complexity classes such as P, NP~\cite{Papadimitriou:1994vk}, and its main concern is,  given an input program, how efficiently it can be computed. On the other hand, algorithmic complexity deals with a problem of what the most efficient encoding of an input program is. This complementary problem to computation complexity has not yet received enough attention 
in quantum information science,
and it would require a further work on quantum version of Kolmogorov complexity~\cite{Mueller:2007wc}.

We would like to stress that our analysis of the experimental data is purely
and consistently algorithmic. We do not resort to statistical methods that are alien to the concept of computation. If this approach can be extended to all quantum experiments, it would allow us to bypass the commonly used statistical interpretation of quantum theory.
 
\section{Acknowledgments}
We acknowledge the support of this work by the National Research Foundation \&
Ministry of Education in Singapore, partly through the Academic Research Fund MOE2012-T3-1-009. P.K. and D.K. are also supported by the Foundational Questions Institute (FQXi).
A.C. also thanks Andrea Baronchelli for the hints on the use of compression software.

\section{Appendix: Symmetrization of detector efficiencies}\label{app:symmetrization}

The experimental setup (Fig.~\ref{fig:setup}) uses four APDs:
$D_{HA}$, $D_{VA}$ (Alice), and $D_{HB}$, $D_{VB}$ (Bob) to register photon pair events
in the two spatial modes. By denoting events at $D_{H}$ and $D_{V}$ as 1 and
0, the four possible output patterns are 00, 01, 10, and 11, where the least
and most significant bit corresponds to the Alice and Bob mode,
respectively. Due to differences in the the losses in the transmitted and
reflected port of the PBS, efficiencies in coupling light into the APDs, and
the quantum efficiencies of APDs, the detection efficiencies for the four
output combinations are different. 
The resulting effective pair efficiencies are then given by the product of the
contributing detection efficiencies $\eta_{VB}$, $\eta_{HB}$, $\eta_{VA}$, and $\eta_{HA}$.

This asymmetry will skew the statistics of the measurement results.
We symmetrize the effective pair efficiencies for each  $(\theta_{A}, \theta_{B})$
measuring also the following settings: 
$(\theta_{A}+45^\circ,\theta_{B})$,
$(\theta_{A}, \theta_{B}+45^\circ)$,
and $(\theta_{A}+45^\circ,\theta_{B}+45^\circ)$.
This procedure
swaps the output ports of
the PBS at which each
polarization state is detected.
The resulting outcomes are then interleaved,
providing an uniform detection
probability
for the four possible outcomes.
The effective pair detection efficiency for all four combinations is then
$(\eta_{VB}\,\eta_{VA}\,+\,\eta_{VB}\,\eta_{HA}\,+\,\eta_{HB}\,\eta_{VA}\,+\,\eta_{HB}\,\eta_{HA})/4$.

\section{Appendix: Choice of compressor}\label{app:choice_compression}
In order to evaluate the NCDs of the binary strings, we need to choose a
compression algorithm that performs close to the Shannon limit. This is
necessary to ensure that it does not introduce any unintended
artifacts that lead to an overestimation of the violation. Preferably we
want to work in the regime where the obtained NCDs
always underestimate the violation. For this purpose, we characterized
four compression algorithms implemented by freely available compression
programs:
\textit{lzma}~\cite{ref:lzma}, \textit{bzip2}~\cite{ref:bzip2}, \textit{gzip}~\cite{ref:gzip},
and \textit{lzw}~\cite{Welch:84}. To eliminate
the overhead associated with the compression of ASCII text files, we
save data in a binary format.

For this characterization and a simulation of the experiment, we need to
generate a ``random'' string of bits (1, 0) or pairs of bits (00, 01, 10, and
11) of various length with various probability distributions. 
We generate these strings using the \emph{MATLAB}~\cite{ref:matlab} function
\mbox{\emph{randsample()}} that uses the pseudo random number generator
\emph{mt19937ar} with a long period of $2^{19937} - 1$. It is based on the
Mersenne Twister~\cite{matsumoto:98}, with ziggurat~\cite{marsaglia:00} as the
algorithm that generates the required probability distribution. The complexity
of this (deterministic) source of pseudorandom numbers should be high enough
to {\emph not} be captured as algorithmic.

\begin{figure}
  \begin{center}
    \includegraphics[width = 0.95\columnwidth]{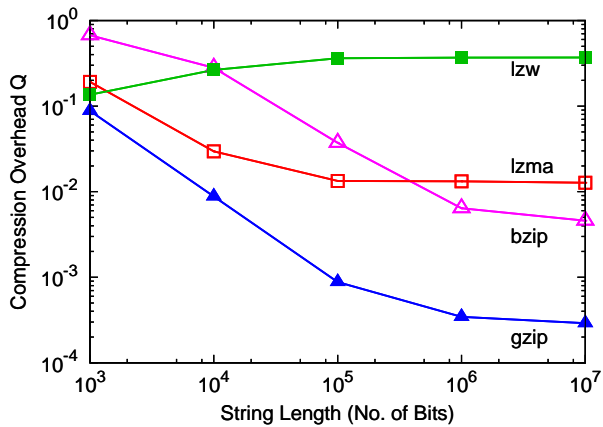}
        \caption{\label{fig:compression_ratio_singles}
        Comparison of the compression overhead $Q$ obtained using four different compression algorithms on pseudo-random strings of varying lengths. The expected value for an ideal compressor is 0.
         From this characterization we can exclude \textit{lzw} as a useful compressor for our application.
        }
    \end{center}
\end{figure}

The first part of this characterization involves establishing the minimum string length required for the compression algorithms to perform consistently.
We start by generating binary strings, $x$, with equal probability of 1's and 0's, i.e. random strings, of varying length.
For each $x$, we evaluate the compression overhead $Q$ as
\begin{equation}\label{eq:compression_overhead}
    Q=\frac{C(x)-H(x)} {l(x)}\,.
\end{equation}
For a good compressor, we expect $Q$ to be close to 0.
From Fig.~\ref{fig:compression_ratio_singles}, it can be seen
that for all the compressors, $Q$ starts to converge after about
$10^5$ bits, setting the minimum string length required for the compressors to
work consistently.
The \textit{lzw} compressor fails this test, converging to
a $Q$ of 0.37 for long string, 
while \textit{bzip2}, \textit{gzip}, and \textit{lzma} give a $Q$ 
below $10^{-1}$.

In the second part of this characterization, 
test the compressors with strings with a known amount of correlation.
We generate a random string $x$ of length
$10^7$ using the same technique already described.
We then generate a second string $y$ of equal length 
and with probability $p$ of being correlated
to $x$.
For $p=0$ the two strings are equal, i.e. perfectly correlated.
For $p=0.5$ they are uncorrelated.

\begin{figure}
  \begin{center}
 \includegraphics[width = 0.95\columnwidth]{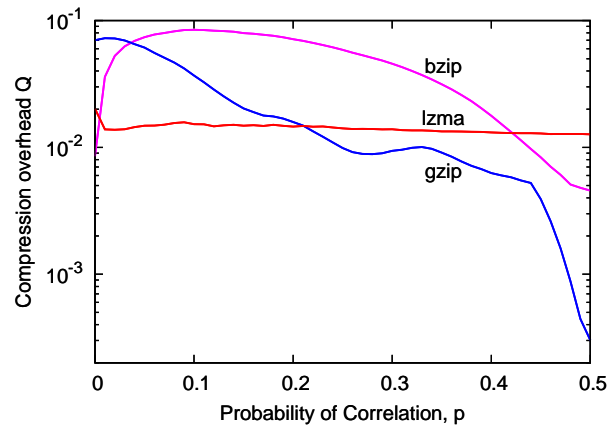}
        \caption{
          Compression overhead $Q$ for the string $xy$ as a function of the probability of pairwise correlation $p$ between the bits of the generating strings $x$ and $y$
          for three different compressors: \textit{bzip}, \textit{gzip}, and \textit{lzma}.
        \label{fig:compressibility_versus_correlation}}
    \end{center}
\end{figure}

The two strings $x$ and $y$ are then combined to form the string $xy$:
to avoid artifacts due to the limited data block size of the compression
algorithms, the elements of $x$ and $y$ are interleaved.
We then compress $xy$ and evaluate the compression overhead $Q$ as a function of $p$.
The results for different compressors are shown in
Fig.~\ref{fig:compressibility_versus_correlation}. Although 
there are ranges of $p$
where \textit{bzip} and \textit{gzip} perform better than \textit{lzma}, the latter shows a more uniform performance over the entire interval of $p$.
It is reasonable to assume that 
the use of \textit{lzma}
should reduce the possibility of artifacts in the estimation of the NCD also for the data obtained from the experiment.


\bibliographystyle{apsrev4-1}
\bibliography{compression.bib}{}

\end{document}